\documentclass [letter,12pt, twoside] {report}
\usepackage[dvips]{graphicx}
\usepackage{amsfonts,amsmath,amssymb,amscd,array,euscript}

\makeatletter

\renewcommand{\section}{\@startsection{section}{1}%
{\parindent}{3.5ex plus 1ex minus .2 ex} {1.5 ex plus .2 ex}{\large\bf}}

\renewcommand{\subsection}{\@startsection{subsection}{1}%
{\parindent}{3.5ex plus 1ex minus .2 ex} {.5 ex plus .2 ex}{\normalsize\bf}}

\renewcommand{\sectionmark}[1]{}

\newcommand{\l@abcd}[2]{#1\dotfill #2 \\}

\makeatother

\newcommand {\bge} {\begin{equation}}
\newcommand {\ee} {\end{equation}}
\newcommand {\bgen} {\begin{equation*}}
\newcommand {\een} {\end{equation*}}
\newcommand {\bgml} {\begin{multline}}
\newcommand {\eml} {\end{multline}}
\newcommand {\bgmln} {\begin{multline*}}
\newcommand {\emln} {\end{multline*}}   % не работает
\newcommand {\bgg} {\begin{gather}}
\newcommand {\eg} {\end{gather}}     % не работает
\newcommand {\bga} {\begin{array}}
\newcommand {\ea} {\end{array}}
\newcommand {\bgp} {\begin{picture}}
\newcommand {\ep} {\end{picture}}
\newcommand {\bgc} {\begin{center}}
\newcommand {\ec} {\end{center}}
\newcommand {\bgt} {\begin{tabular}}
\newcommand {\et} {\end{tabular}}

\newcommand {\nin}{\noindent}

\newcommand {\mes} {\medskip}

\newcommand{\bea}{\begin{eqnarray}}
\newcommand{\eea}{\end{eqnarray}}

\def\2#1#2#3{{#1}_{#2}\hspace{0pt}^{#3}}
\def\3#1#2#3#4{{#1}_{#2}\hspace{0pt}^{#3}\hspace{0pt}_{#4}}
\newcounter{sctn}
\def\sec#1.#2\par{\setcounter{sctn}{#1}\setcounter{equation}{0}
                  \noindent{\bf\boldmath#1.#2}\bigskip\par}

\def\thebibliography#1{\subsubsection*{References}
\list
  {[\arabic{enumi}]}{\settowidth\labelwidth{[#1]}\leftmargin\labelwidth
  \advance\leftmargin\labelsep
  \usecounter{enumi}}
  \def\newblock{\hskip .11em plus 0.33em minus -.07em}
  \sloppy
  \sfcode`\.=1000\relax}
 
 \catcode `\@=12
\newcommand {\etit}[1] {\bgc{\textbf{\large{#1}}} \mes\mes\ec}  % E-Title
\newcommand {\eauth}[2] {\bgc{\textbf{\large{#1}}\\ \mes\mes \emph{#2}} \mes\mes\ec}  % E-Author
\newcommand {\esect}[1] {\mes\mes\mes\mes\nin{\textbf{#1}} \mes\mes}  % E-Section

 \newcommand {\start} {{\setcounter{equation}{0}}
                     \setcounter{footnote}{0} \setcounter{figure}{0} \setcounter{table}{0}}
\begin{document}

\start

\etit{GENERALIZATION OF\\ \mes CONFORMAL TRANSFORMATIONS}

\eauth{G. I. Garas'ko} {Russian Institute for Electrotechnics, Moscow, Russia\\
gri9z@mail.ru}

%\address

{\small Conformal transformations of a Euclidean (complex) plane have
    some kind of completeness (sufficiency) for the solution of many
    mathematical and physical-mathematical problems formulated on
    this plane. There is no such completeness in the case of
    Euclidean, pseudo-Euclidean and polynumber spaces of dimension
    greater than two. In the present paper we show that using the
    concepts of analogical geometries allows us to generalize
    conformal transformations not only to the case of Euclidean or
    pseudo-Euclidean spaces, but also to the case of Finsler spaces,
    analogous to the spaces of affine connectedness. Examples of such
    transformations in the case of complex and hypercomplex
    numbers $H_4$ are presented. In the general case such transformations
    form a group of transitions, the elements of which can be viewed
    as transitions between projective Euclidean geometries of a
    distinguished class fixed by the choice of metric geometry admitting
    affine coordinates. The correlation between functions realizing
    generalized conformal transformations and generalized analytical
    functions can appear to be productive for the solution of
    fundamental problems in theoretical and mathematical physics.}

\esect{Introduction}

\qquad Conformal transformations play a distinguished role in mathematics and
physics. Riemanian and pseudo-Riemanian spaces of constant curvature are not
less important (among such spaces are Lobachevsky space and spherical space),
their homogeneity is as complete as in the case of a Euclidean space, since
their motion groups have the same number of parameters as in the Euclidean case
\cite{1}. This work studies only Finslerian spaces admitting an affine
coordinate system, so in the case of metric spaces, that is why we consider the
length element to be the basic concept, and the concept of angle will be
considered secondary. The proposed approach (of course, changed slightly) can
be also applied for spaces (geometries) having the length element not defined,
but with angles between vectors defined in each point.

If $V_n$ is a Riemanian or a pseudo-Riemanian space with coordinates $x^i$ and
a metric tensor $g_{ij}(x)$, then the connection coefficients $\Gamma^i_{kl}$
in this space are well-known to be defined by the following formula:
\begin{equation}\label{1}
\Gamma^i_{kl}(g)=\frac{1}{2} g^{im}\left( \frac{\partial g_{mk}
}{\partial x^l}+\frac{\partial g_{ml} }{\partial
x^k}-\frac{\partial g_{kl}}{\partial x^m} \right) .
\end{equation}
If
\begin{equation}\label{2}
G_{ij}(x)=\Lambda(x)\cdot g_{ij}(x),
\end{equation}
where $\Lambda(x)>0$ is a scalar function defined on coordinates,
then
\begin{equation}\label{3}
\Gamma^i_{kl}(G)=\Gamma^i_{kl}(g)+\frac{1}{2\Lambda} \left(
\frac{\partial \Lambda }{\partial x^l}\delta^i_k+\frac{\partial
\Lambda}{\partial x^k}\delta^i_l-g^{im}\frac{\partial
\Lambda}{\partial x^m} g_{kl} \right) .
\end{equation}
Spaces with metric tensors $g_{ij}$ and $G_{ij}$ are called conformally
connected \cite{1}.

Since connectivity coefficients are transformed by the following formulas when
changing the coordinate system:
\begin{equation}\label{4}
\frac{\partial x^{i'}}{\partial x^i} \Gamma^i_{k l} = \Gamma^{i'}_{n' p'}
\frac{\partial x^{n'}}{\partial x^k}\frac{\partial x^{p'}}{\partial x^l} +
\frac{\partial^2 x^{i'}}{\partial x^k \partial x^l}.
\end{equation}
These are conformal transformations of coordinates, realized by functions $f^i$
in some area $W_n\subset V_n$, where the metric tensor $g_{ij}$ does not depend
on the point of the space, and they satisfy the following system of equations:
\begin{equation}\label{5}
\frac{\partial^2 f^i}{\partial x^k \partial x^l} =
\frac{1}{2\Lambda} \left( \frac{\partial \Lambda }{\partial
x^l}\delta^m_k+\frac{\partial \Lambda}{\partial
x^k}\delta^m_l-g^{mp}\frac{\partial \Lambda}{\partial x^p} g_{kl}
\right)\frac{\partial f^i}{\partial x^m}.
\end{equation}

The convolution of both sides of the equations (\ref{5}) and the tensor
$g^{kl}$ over both indexes gives us the following:
\begin{equation}\label{6}
g^{kl} \frac{\partial^2 f^i}{\partial x^k \partial x^l} =
\frac{2-n}{2\Lambda} g^{kl} \frac{\partial \Lambda}{\partial x^k}
 \frac{\partial f^i}{\partial x^l}.
\end{equation}
Thus the functions realizing conformal transformation in Euclidean and
pseudo-Euclidean spaces are the solutions of the differential equation
(\ref{6}).

For analytical functions of a complex variable (the first type conformal
transformations of the Euclidean plane) and for complex conjugate analytical
function of a complex variable (the second type conformal transformations of
the Euclidean plane)

\begin{equation}\label{7}
\Lambda = \left( \frac{\partial f^1}{\partial x^1}
\right)^2+\left( \frac{\partial f^1}{\partial x^2} \right)^2,
\end{equation}
and the equations (\ref{5}) are valid in the area of analyticity
and simple-connected\-ness.

\esect{Generalization of conformal transformations \\ in Euclidean and
pseudo-Euclidean spaces}

\qquad The concept of analogical geometries was introduced in
\cite{2}. It is proposed to call geometries analogical in some
areas, if these geometries have same dimen\-sions and if there
exists a mapping of one area onto another, under which some set of
geodesics (extremals) of one geometry is mapped exactly on some
set of geodesics (extremals) of the second geometry. Under certain
assumptions the similarity of geometries means that there exist
coordinate systems in which the differential equations of
geodesics (extremals) coincide.

If in some geometry of affine connectedness we add to the connectivity
coefficients the tensor
\begin{equation}\label{8}
T^i_{kl}=\frac{1}{2}(p_k \delta^i_l + p_l \delta^i_k)+ S^i_{kl} ,
\end{equation}
where $p_i$ is an arbitrary covariant field, and $S^i_{kl}$ is an arbitrary
tensor field, antisymmetric with respect to the lower two indexes, then the
geodesic curves will remain the same \cite{1}.

Let functions $f^i$ map an area in a Euclidean or a pseudo-Euclidean space with
a metric tensor $g_{ij}$ bijectively onto another area in the same space, and
suppose also that these functions satisfy the following system of equations:
\begin{equation}\label{9}
\frac{\partial^2 f^i}{\partial x^k \partial x^l} = \left[
\frac{1}{2}(p_l \delta^m_k+p_k \delta^m_l)-g^{mp}\frac{\partial
L}{\partial x^p} g_{kl} \right] \frac{\partial f^i}{\partial x^m}
\, ,
\end{equation}
where $p_i$ is a covariant vector field and $L$ a scalar field. Then this map
(a coordinate transformation) will be called elementary generalized conformal.

Notice that in the case of using the additional term (\ref{8}) with non-zero
torsion tensor $S^i_{kl}$ to obtain the formulas (\ref{9}) instead of a
generalization additional conditions appear:
\begin{equation}\label{10}
S^m_{kl}\frac{\partial f^i}{\partial x^m} = 0 \,,
\end{equation}
as far as all the other additive terms in both sides of the system (\ref{9})
are symmetric under the permutations of indexes $k$ and $l$.

It follows from the definition of elementary generalized conformal
trans\-for\-ma\-tions of Euclidean and pseudo--Euclidean spaces
that these transformations and functions $f^i$ realizing them are
closely connected with the concept of projec\-tive Euclidean
geometries \cite{1}.

Thus each function (a component) of an elementary generalized conformal
transformation satisfies the following scalar equation:
\begin{equation}\label{11}
g^{kl}\frac{\partial^2 f^i}{\partial x^k \partial x^l} = g^{kl}
\left( p_k -\frac{n}{2} \frac{\partial L}{\partial x^k}
\right)\frac{\partial f^i}{\partial x^l}.
\end{equation}

Though for proper generalized conformal transformations the formula (\ref{2})
is not valid, we will suppose by definition that
\begin{equation}\label{12}
\Lambda = \Lambda_0 \cdot \exp(L).
\end{equation}
In certain sense the scalar field $\Lambda$ defined this way will be a
characteristic for the squared coefficient of the space "stress--strain" under
an elementary generalized conformal transformation.

To show the non--triviality of such a generalization let us perform a solution
of the system (\ref{9}):
\begin{equation}\label{13}
f^i=\frac{x^i}{a+b\cdot g_{kl}x^kx^l},
\end{equation}
where $a$ and $b$ --- are real numbers and
\begin{equation}\label{14}
\Lambda=\frac{d}{(a-b\cdot g_{kl}x^kx^l)^2},
\end{equation}
where $d$ is a real number.

In the case of the Euclidean (complex) plane $(x,y)$
\begin{equation}\label{15}
z=x+iy, \qquad F(z)=f^1+if^2,
\end{equation}
the function (\ref{13})
\begin{equation}\label{16}
F(z) = \frac{z}{a+bz\bar{z}}
\end{equation}
is neither analytical nor complex conjugate analytical when $a\neq 0$ and
$b\neq 0$, but it realizes an elementary generalized conformal transformation
of the plane. When $a=0$ this function becomes complex conjugate analytical
\begin{equation}\label{17}
F(z)=\frac{1}{b\bar{z}},
\end{equation}
which corresponds to a conformal map of the second type. When $b=0$ the
function $F(z)$ is analytical,
\begin{equation}\label{18} F(z)=\frac{1}{a}z,
\end{equation}
which corresponds to a conformal map of the first type.

\esect{Polynumbers $H_4$}

\qquad In the space $H_4$ the fourth power of the length element written in the
basis $\psi$ looks like
\begin{equation}\label{19}
(ds)^4=d\xi^1d\xi^2d\xi^3d\xi^4,
\end{equation}
and a conformally connected geometry will have the length element
\begin{equation}\label{20}
(ds)^4=\Xi d\xi^1d\xi^2d\xi^3d\xi^4,
\end{equation}
where $\Xi>0$ is a scalar field. This geometry is similar to the geometry of
affine connectedness with the connectivity coefficients \cite{2}
\begin{equation}\label{21}
\Gamma^i_{kj}=\frac{1}{2}(p_k\delta^i_j+p_j\delta^i_k)-
p^i_{kj}\frac{1}{\Xi}\frac{\partial \Xi}{\partial
\xi^{j_-}}+S^i_{kj},
\end{equation}
where
\begin{equation}\label{22}
\psi_k \psi_j=p^i_{kj} \psi_i, \qquad
p^i_{kj}=\left\{\begin{array}{l}
  1,\, \hbox{if} \, \, i=j=k, \\
   \\
  0,\, \hbox{otherwise},
\end{array}   \right .
\end{equation}
$p_k, \, S^i_{kj}=-S^i_{jk}$ are arbitrary tensor fields.

Thus we obtain the system of equations for functions $f^i$, which realize an
elementary generalized conformal transformation in the coordinate space of
polynumbers $H_4$:
\begin{equation}\label{23}
\frac{\partial^2 f^i}{\partial \xi^k \partial \xi^l} = \left[
\frac{1}{2}(p_l \delta^m_k+p_k \delta^m_l)-p^m_{kl}\frac{\partial
L}{\partial \xi^{l_-}} \right] \frac{\partial f^i}{\partial \xi^m},
\end{equation}
where
\begin{equation}\label{24}
\Xi = \Xi_0 \cdot \exp(-L).
\end{equation}

Any function analytical with respect to the variable $H_4$ realizing a
one-to-one correspondence between two ares contained in the coordinate space of
polynumbers $H_4$ satisfies the system (\ref{23}), and at the same time
\begin{equation}\label{25}
p_i=0, \qquad \Xi=\dot{f}^1\dot{f}^2\dot{f}^3\dot{f}^4, \qquad
L=-\ln |\Xi /\Xi_0|.
\end{equation}
Functions analytical with respect to the variable $H_4$ are not the only
solutions of the system (\ref{23}). Another solution of this system is the
function
\begin{equation}\label{26}
f^i=\frac{f^i_0 \ln \left| \displaystyle
\frac{\xi^{i_-}}{\xi^{i_-}_0} \right|}{a+b \ln \left|
\displaystyle\frac{\xi^1\xi^2\xi^3\xi^4}{\xi^1_0\xi^2_0\xi^3_0\xi^4_0}
\right|} ,
\end{equation}
which becomes analytical with respect to the variable $H_4$ only when $b=0$. In
the formula (\ref{26}) $a, \, b, \, \xi^i_0, \, f^i_0 $ are constants but, of
course, they are not all independent. For the function (\ref{26})
\begin{equation}\label{27}
\Xi=\frac{const}{\xi^1\xi^2\xi^3\xi^4} .
\end{equation}

As far as in the space $H_4$ the following tensor can be defined
\begin{equation}\label{28}
q_{ij}=p^m_{ik}p^k_{mj}, \qquad (q_{ij})=diag(1,1,1,1) ,
\end{equation}
there also exists a twice contravariant tensor $q^{ij}$,
\begin{equation}\label{29}
(q^{ij})=diag(1,1,1,1) .
\end{equation}
This is why each component of an elementary generalized conformal
trans\-for\-ma\-tion of $H_4$ should satisfy the following scalar
equation:
\begin{equation}\label{30}
q^{kl}\frac{\partial^2 f^i}{\partial \xi^k \partial \xi^l} = q^{kl}
\left( p_k-\frac{\partial L}{\partial \xi^k}\right) \frac{\partial
f^i}{\partial \xi^l}.
\end{equation}
Comparing the equations (\ref{11}), (\ref{30}) and taking into
account the formulas (\ref{12}) (\ref{24}), we see that the scalar
equation (\ref{11}), solutions of which are the functions
realizing generalized conformal transformations in the
four--dimensional Eucli\-dean space, and the scalar equation
(\ref{30}) describing the functions realizing generalized
conformal transformation in the space $H_4$ have the same
structure:
\begin{equation}\label{31}
\delta^{kl}\frac{\partial^2 f^i}{\partial \xi^k \partial \xi^l}
=\delta^{kl}\left( p_k \mp 4 \frac{\partial l}{\partial
\xi^k}\right) \frac{\partial f^i}{\partial \xi^l} ,
\end{equation}
where the coefficient $\lambda$ of linear "stress--strain"\ can be expressed in
the terms of a scalar field $l$ for the both four--dimensional Euclidean space
and space $H_4$ with the same formula
\begin{equation}\label{32}
\lambda=\lambda_0\exp(l) .
\end{equation}
Notice, however, that we cannot claim that  $p_k$ and $l$ are the
same in the four--dimensional Euclidean space and in the space
$H_4$. At the same time it would be very interesting to find such a
class of elementary generalized conformal transformations, that for
all its elements the covariant field $\left( p_k - 4 \displaystyle
\frac{\partial l}{\partial \xi^k} \right)_{E_4} = \left( p_k + 4
\displaystyle \frac{\partial l}{\partial \xi^k} \right)_{H_4}$ would
be the same in the four--dimensional Euclidean space and in the
space $H_4$, i.e. that in both cases the functions $f^i$ would
satisfy the same scalar equation not only formally. Linear
transforms automatically form a subset of such a class of
transformations.

\esect{Generalized conformal transformations}

\qquad The preceding constructions allow us to suppose that the
system of equa\-tions defining elementary generalized conformal
transformations of a metric geometry (at this moment Finsler
geometry is developed more than enough for the needs of
theoretical and mathematical physics) admitting affine coordinates
and for which all its conformally connected spaces are always
similar to some geometry of affine connectedness has the following
most general view in the affine coordinates:

\begin{equation}\label{33}
\frac{\partial^2 f^i}{\partial x^k \partial x^l} = \left[
\frac{1}{2}(p_l \delta^m_k+p_k
\delta^m_l)-\Delta^{pm}_{kl}\frac{\partial L}{\partial x^p}
\right] \frac{\partial f^i}{\partial x^m},
\end{equation}
where $\Delta^{pm}_{kl}$ is a symmetric with respect to the lower indexes
number tensor in an affine coordinate system of the initial metric geometry,
$L$ and $p_k$ are a scalar and a covariant fields; and for conformal transforms
the coefficient $\lambda$ of linear "stress--strain"\ is expressed in the terms
of the scalar field $L$ with the formula
\begin{equation}\label{34}
\lambda=\lambda_0\exp(\pm L/m)\equiv \lambda_0\exp(l)  .
\end{equation}
Here $\lambda_0$ is a real number and $m$ is a natural number, equal to the
order of the Finsler geometry form, by which the length element is expressed,
for instance, for Euclidean and pseudo--Euclidean geometries $m=2$ and for
$H_4$--numbers $m=4$.

It follows from the formulas (\ref{33}) that any linear
non--degenerate trans\-for\-ma\-tion is elementary generalized
conformal with
\begin{equation}\label{35}
p_i=0, \qquad L=const .
\end{equation}

Though we do hope that for all possible tensors $\Delta^{pm}_{kl}$ the concept
of Finsler geometry is enough (it is possible that the concept of polynomial
geometry \cite{3} might be enough), this conjecture (same as the stronger one)
needs a rigorous proof.

For non--degenerate polynumber spaces $P_n$ there always exists a tensor
$q^{ij}$ (see (\ref{28}), (\ref{29})), that is why in such spaces elementary
generalized conformal transformations satisfy the following scalar equation:
\begin{equation}\label{36}
q^{kl} \frac{\partial^2 f^i}{\partial x^k \partial x^l} = \left(
p_k q^{km} - q^{kl} \Delta^{pm}_{kl}\frac{\partial L}{\partial
x^p} \right) \frac{\partial f^i}{\partial x^m} .
\end{equation}

Elementary generalized conformal transformations (\ref{33}) do not form a group. But
all their products (i.e. consequent executions) together with the inverse ones do
form a group, which will be denoted as $G_n(\Delta^{pm}_{kl})$ and called a group of
generalized conformal transformations. Products of elementary generalized conformal
transformation with the inverse of another one are the solutions of the system
\begin{equation}\label{37}
\begin{array}{c} \displaystyle{ \frac{\partial^2 f^i}{\partial x^k
\partial x^l} =  \left[ \frac{1}{2}(p_l \delta^m_k+p_k
\delta^m_l)-\Delta^{pm}_{kl} \frac{\partial L}{\partial x^p}
\right] \frac{\partial f^i}{\partial x^m} \, -  } \\
   \\
\displaystyle{\qquad \qquad  -  \left[ \frac{1}{2}(p'_r \delta^i_s
+ p'_s \delta^i_r)-\Delta^{pi}_{sr}\frac{\partial L'}{\partial
f^p} \right] \frac{\partial f^s}{\partial x^k} \frac{\partial
f^r}{\partial x^l} ,}
\end{array}
\end{equation}
where $p_l, \, p'_k, \, L, \, L'$ are some fields, $\Delta^{pi}_{sr}$ is the
same scalar tensor as in the system of equations (\ref{33}); and the
derivatives $\displaystyle\frac{\partial L}{\partial f^p}$ are meant to be
explicitly expressed it terms of partial derivatives by $x^i$.

Generalized conformal transformations can be viewed as transitions in the uniquely
characterized by the tensor $\Delta^{pi}_{sr}$ subset (class) of projective
Euclidean spaces. Let us emphasize once again that it is enough to investigate
elementary generalized conformal transformations, because an arbitrary generalized
con\-for\-mal transformation can be constructed as a product of elementary
trans\-for\-ma\-tions and inverse to elementary transformations.

\esect{Generalized analytical functions}

\qquad If the initial metric space with a number tensor $\Delta^{pm}_{kl}$
corresponding to it is polynumber $P_n\ni X$, then analytical functions realize
conformal transformations in the area where the Jacobean of their coordinates
is different from zero, and a concept of generalized analytical functions can
be introduced in this space \cite{3}. Of course, in this case functions
realizing generalized conformal transformations are generalized analytical
functions of the given polynumber variable. The following problem seems to be
more interesting: find a class $\Upsilon(\Delta^{pm}_{kl})\ni F(X)$ of
generalized analytical functions, each element of which is a solution of the
system (\ref{37}).

Notice that if $F_{(1)}(X),\,F_{(2)}\in\Upsilon(\Delta^{pm}_{kl})$, then
$F_{(1)}\left(F_{(2)}\right)\in\Upsilon(\Delta^{pm}_{kl})$. It follows from the
group properties of generalized conformal transformations.

A generalized analytical function of a polynumber variable $X \in P_n$,
\begin{equation}\label{38}
F(X)=f^1(x^1,x^2,...,x^n)e_1 + f^2(x^1,x^2,...,x^n)e_2 + ... +
f^n(x^1,x^2,...,x^n)e_n ,
\end{equation}
$X = x^ie_i$, $e_i$ is a basis, satisfies the correlations
\begin{equation}\label{39}
\frac{\partial f^i}{\partial x^k}+\gamma^i_k = p^i_{kj}\dot{f}^j ,
\end{equation}
where $\dot{f}^j$ is a generalized derivative, tensor $p^i_{kj}$ is defined by
the correlations
\begin{equation}\label{40}
e_ke_j = p^i_{kj} e_i ,
\end{equation}
and the object $\gamma^i_k$ should change under transition to another
coordinate system according to the following law
\begin{equation}\label{41}
\gamma^{i'}_{k'} = \frac{\partial x^k}{\partial x^{k'}}
\frac{\partial x^{i'}}{\partial x^i}\gamma^i_k - \frac{\partial
x^k}{\partial x^{k'}} \frac{\partial^2 x^{i'}}{\partial x^k
\partial x^i} f^i .
\end{equation}
If $\varepsilon^i$ are the coefficients of the unit's decomposition in the
basis $e_i$ then taking into account the following formula:
\begin{equation}\label{42}
\varepsilon^kp^i_{kj} = \delta^i_j ,
\end{equation}
from the formula (\ref{39}) we get
\begin{equation}\label{43}
\dot{f}^i = \varepsilon^m \frac{\partial f^i}{\partial x^m}+
\varepsilon^m \gamma^i_m
\end{equation}
and an analogue of Cauchy--Riemann correlations:
\begin{equation}\label{44}
\frac{\partial f^i}{\partial x^k}+\gamma^i_k - p^i_{kj} \left(
\varepsilon^m \frac{\partial f^j}{\partial x^m}+ \varepsilon^m
\gamma^j_m \right) = 0 .
\end{equation}
The conditions of correlations (\ref{39}) integrability (with respect to the
functions $f^i$) are as follows:
\begin{equation}\label{45}
\frac{\partial}{\partial x^m}\left(- \gamma^i_k +
p^i_{kj}\dot{f}^j \right) = \frac{\partial}{\partial
x^k}\left(-\gamma^i_m + p^i_{mj}\dot{f}^j \right) .
\end{equation}

If the polynumbers system $P_n$ is non--degenerate and the
generalized deriva\-tive is also a generalized analytical function
$\{\dot{f}^i,\dot{\gamma}^i_k \}$ then each component $f^i$
formally satisfies the following scalar equation:
\begin{equation}\label{46}
q^{mk}\dot{\tilde{\nabla}}_m\tilde{\nabla}_kf^i=Q^i_r\ddot{f}^r,
\end{equation}
where
\begin{equation}\label{47}
Q^i_r=q^{mk}p^i_{kj}p^j_{mr}.
\end{equation}
For analytical functions of a complex variable this equation becomes
\begin{equation}\label{48}
\left( \frac{\partial^2}{\partial x^2} -
\frac{\partial^2}{\partial y ^2}  \right) f^i = 2\ddot{f}^i
\end{equation}
and is identical. Thus the field $(2\ddot{f}^i)$ can be considered as the field
of a field source $f^i$ for the operator
\begin{equation}\label{49}
\frac{\partial^2}{\partial x^2} - \frac{\partial^2}{\partial y^2}.
\end{equation}

Consider a two--dimensional non--homogeneous (with the right--hand side)
hyperbolic equation in partial derivatives.
\begin{equation}\label{50}
\left( \frac{\partial^2}{\partial t^2} - a^2
\frac{\partial^2}{\partial x^2_s} \right) u_s = f_s(t, x_s),
\end{equation}
where $t$ is time, $x_s$ is the coordinate along the string, $u_s(t,x_s)$ is
the amplitude of small lateral oscillations of the string, $ \rho f_s dx_s$ is
the lateral force acting on an element $(x_s,x_s+dx_s)$ of the string, $\rho$
is the mass density. When changing the variables
\begin{equation}\label{51}
f^i=u_s, \qquad  at=x,   \qquad  y=x_s \, , \qquad
\frac{1}{a^2}f_s(t,x_s)=2f^i(x,y) \qquad
\end{equation}
the equations (\ref{48}) and (\ref{50}) switch places but the
right--hand side of the equation (\ref{48}) is an analytical
function of a complex variable $(x,y)$, which restricts
suf\-ficient\-ly the variety of sources.

Thus if the source function (the right--hand side) of a two--dimensional
non--homogeneous hyperbolic equation (a wave equation) written in a special
form (\ref{48}) is an analytical function of a complex variable, then one of
the solutions of this equation will be the second antiderivative of the source
function divided by two.

Except for the equation (\ref{48}) each analytical function of a complex
variable satisfies the Laplace equation, which can be obtained analogically to
how the equation (\ref{46}) was obtained, having changed the tensor $q^{mk}$
into the tensor $g^{mk}$, which is inverse to $g_{ij}$, the metric tensor of
the Euclidean plane:
\begin{equation}\label{52}
g^{mk}\dot{\tilde{\nabla}}_m\tilde{\nabla}_kf^i=0  \Rightarrow
\left( \frac{\partial^2}{\partial x^2} +\frac{\partial^2}{\partial
y ^2}  \right)f^i = 0 .
\end{equation}

Similar equations are valid also for analytical functions of an
$H_2$--variable,
\begin{equation}\label{53}
X=x+jy, \qquad j^2=1,
\end{equation}
but the elliptic and hyperbolic types of equations switch places:
\begin{equation}\label{54}
\left( \frac{\partial^2}{\partial x^2} +
\frac{\partial^2}{\partial y ^2}  \right)f^i = 2\ddot{f}^i, \quad
\left( \frac{\partial^2}{\partial x^2} -
\frac{\partial^2}{\partial y ^2} \right)f^i = 0.
\end{equation}

So, if the source function (the right--hand side) of a two--dimensional
non--homogeneous Laplace equation is an analytical function of an
$H_2$--variable, then one of the solutions of this equation will be the second
antiderivative of the source function divided by two.

Thus when changing $C \leftrightarrow H_2$ not only the wave equation and
Laplace equation "switch places", but one of them loses the source (the
non--homogeneous right side) and another one gains it. It is quite reasonable
now to suppose that such symmetry might take place for polynumbers of dimension
greater than two and not only for analytical but also for generalized
analytical functions.

The scalar equation (\ref{46}) for analytical functions of an $H_4$--variable
in the coordinate system of the $\psi$--basis (\ref{22}) becomes
\begin{equation}\label{55}
\left( \frac{\partial^2}{\partial (\xi^1)^2} +
\frac{\partial^2}{\partial (\xi ^2)^2} +
\frac{\partial^2}{\partial (\xi^3)^2} + \frac{\partial^2}{\partial
(\xi ^4)^2} \right)f^i = \ddot{f}^i,
\end{equation}
or in the coordinate system $(x^0,x^1,x^2,x^3)$ of the basis $\{ 1, \, j, \, k,
\, jk  \}$ consisting of the unit and three symbol units $j^2=k^2=(jk)^2=1$:
\begin{equation}\label{56}
\left. \begin{array}{cc}
  \xi^1=x^0+x^1+x^2+x^3, & \xi^2=x^0+x^1-x^2-x^3, \\
   &  \\
  \xi^3=x^0-x^1+x^2-x^3, & \xi^4=x^0-x^1-x^2+x^3
\end{array}\right\}
\end{equation}
that same equation becomes
\begin{equation}\label{57}
\left( \frac{\partial^2}{\partial (x^0)^2} +
\frac{\partial^2}{\partial (x^1)^2} + \frac{\partial^2}{\partial
(x^2)^2} + \frac{\partial^2}{\partial (x^3)^2} \right)f^i = 4
\ddot{f}^i.
\end{equation}

Thus if the source function (the right--hand side) of a four--dimensional
non--homogeneous Laplace equation is an analytical function of an
$H_4$--variable, then one of the solutions of this equation will be the second
antiderivative of the source function divided by four.

Notice also that in the equations (\ref{48}), the first one in (\ref{54}) and
(\ref{57}) one can take an arbitrary linear combination of the source
function's components and not change the coordinates, because the index $i$ is
free in both sides, and also use the symmetry (which the corresponding
polynumbers do not have) of the scalar operators from the right--hand side to
change the coordinates not "shuffling"\ the components of analytical functions.
These circumstances extend in a way the corresponding set of source functions.

\clearpage

\esect{Conclusion}

\qquad In the present paper a generalization of conformal
transformations of a metric space is proposed. If we restrict
ourselves to considering the spaces admitting affine coordinates
then generalized conformal transformations of a given metric space
can be considered as the group of transitions between the elements
of some class of spaces of constant curvature \cite{1}.

If the problem of finding a one--to--one correspondence (modulo a discrete
group of transformations) between generalized conformal transformations of the
space $P_n$ and generalized analytical functions of the polynumber variable
$P_n$ is solved then it is reasonable to hope to build a powerful mathematical
instrument for solving mathematical problems and problems of theoretical
physics appearing in the spaces $P_n$.

\end{document}